\documentclass[12pt,preprint]{aastex}

\usepackage{epsfig}

\shorttitle{X-ray sources in NGC~4552 (M89)}

\shortauthors{Xu et al.}

\begin{document}

\title{{\it Chandra} Study of X-Ray Point Sources in the Early-Type
Galaxy NGC 4552 (M89)}

\author{
Yueheng Xu\altaffilmark{1},
Haiguang Xu\altaffilmark{1,2},
Zhongli Zhang\altaffilmark{1},
Arunav Kundu\altaffilmark{3},
Yu Wang\altaffilmark{4},
and
Xiang-Ping Wu\altaffilmark{2}
}

\altaffiltext{1}{Department of Physics, Shanghai Jiao Tong University, 1954 Huashan Road, Shanghai 200030, P. R. China;
e-mail: yx12@leicester.ac.uk, hgxu@sjtu.edu.cn, zebrafish@sjtu.edu.cn}

\altaffiltext{2}{National Astronomical Observatories, Chinese Academy of Science, 20A Datun Road, Beijing 100012, P. R. China; e-mail: wxp@bao.ac.cn}

\altaffiltext{3}{Department of Physics and Astronomy, Michigan State University, East Lansing, MI 48824, USA; akundu@pa.msu.edu}

\altaffiltext{4}{Department of Astronomy, Beijing Normal University, 19 Xinjiekouwai Street, Beijing 100875, P. R. China; wy@bao.ac.cn}

\begin{abstract}
We present a {\it Chandra} ACIS study of the early-type galaxy NGC 4552.
We detect 47 X-ray point sources, most of which are likely low mass X-ray
binaries (LMXBs), within 4 effective radii ($R_{\rm e}$). The brightest
X-ray source coincides with the optical, UV and radio center of the galaxy,
and shows variability on $>1$ hr timescales, indicating the possible
existence of a low-luminosity AGN. The 46 off-center sources and the
unresolved point sources contribute about 29\% and 20\% to the total luminosity
of the galaxy, respectively. We find that, after correcting for the incompleteness
at the low luminosity end, the observed cumulative X-ray luminosity function
(XLF) of the off-center sources is best fit by a broken power-law model
with a break at
$L_{\rm b} = 4.4^{+2.0}_{-1.4} \times 10^{38}$ ergs s$^{-1}$.
We identified 210 globular clusters (GCs) candidates in a HST WFPC2 optical
image of the galaxy's central region. Of the 25 off-center LMXBs that fall within
the WFPC2 field of view, 10 sources are coincident with a GC. Thus the fraction
of the GCs hosting bright LMXBs and the fraction of the LMXBs associated with GCs
are 4.8\% and 40\%, respectively. In the $V$ and $I$ bands, the GCs hosting bright
LMXBs are typically 1--2 magnitudes brighter than the GCs with no detected LMXBs.
There are about $1.9 \pm 0.4$ times as many LMXBs in the red, metal-rich GCs as
there are in the blue, metal-poor ones. We find no obvious difference between
the luminosity distributions of LMXBs in GCs and in field, but the cumulative
spectrum of the LMXBs in GCs tends to be softer than that of the LMXBs in field.
We detected three X-ray sources that have isotropic luminosities larger than
$10^{39}$ erg s$^{-1}$. The one located in the joint {\it Chandra}-{\it HST}
field is found to be associated with a GC. By studying the ACIS spectra we
infer that the this may be a candidate black hole system with a mass of
15--135 $M_{\odot}$. One of the other sources with a luminosity brighter than
$10^{39}$ erg s$^{-1}$ reveals temporal variations in brightness on timescales
greater than an hour.
\end{abstract}


\section{INTRODUCTION}
Elliptical and S0 galaxies are luminous sources in the X-ray sky
(e.g., Forman et al. 1985). In terms of their X-ray-to-optical luminosity ratios,
these early-type galaxies can generally be divided into two categories: X-ray bright
galaxies and X-ray faint galaxies. Both X-ray faint and X-ray bright galaxies reveal
a hard X-ray component whose intensity is roughly proportional to the optical
luminosity of the galaxy. In X-ray bright early-type galaxies, the hard spectral
component is often overwhelmed by the emission of the hot diffuse gas. However the
hard X-ray component appears to be significant in X-ray faint galaxies and,
in some cases, dominates the X-ray spectrum. Trinchieri and Fabbiano (1985)
and other later studies of the non-thermal emission ascribed it to the contribution
of a large number of low mass X-ray binaries (LMXBs), such as observed in
the bulges of M31 and our Galaxy (e.g., White et al. 1995). With the advent of
the high spatial resolution observations of the {\it Chandra} X-ray observatory,
hundreds of point sources have been resolved in a number of nearby elliptical
and S0 galaxies and the LMXB-origin has been confirmed (e.g., Sarazin et al. 2000).

Most of the resolved X-ray point sources in early-type galaxies are likely LMXBs
and they may be useful in deciphering the evolutionary history of their host galaxies
and stars therein, and constraining our understanding of the physics of compact stars.
In the past few years {\it Chandra} observations have revealed   that the X-ray luminosities
of the resolved point sources in early type galaxies span a wide range from the typical
observational limit of a few 
$10^{37}$ erg s$^{-1}$ 
to over 
$10^{39}$ erg s$^{-1}$
(e.g., Sarazin et al. 2000; Angelini et al. 2001; Kundu et al. 2002). 
The observed XLFs show quite similar, but not exactly the same profiles among
galaxies. For example, in NGC 4697, Sarazin et al. (2000) reported that there is
a knee on the XLF at around the Eddington luminosity of normal accreting neutron
stars with a mass 
$\simeq 1.4 M_{\rm \odot}$
($L_{\rm Edd} \simeq 2 \times 10^{38}$ ergs s$^{-1}$),
and thus suggested that the break might be an universal feature that can be used 
as a distance indicator. The existence of a similar break was confirmed by Kundu 
et al. (2002) in NGC 4472 and by Blanton et al. (2001) in NGC 1553. However, 
by analyzing a sample of 14 E/S0 galaxies, Kim and Fabbiano (2004) argued that 
the position of the break is significantly higher than $ L_{\rm Edd}$ of normal
neutron stars. In NGC 720, Jeltema et al. (2003) showed an even higher break at
$L_{\rm b} = 2.10^{+0.2}_{-0.2} \times 10^{39}$ ergs s$^{-1}$ (although this
is based on a distance derived from an adopted $H_0$ of 50 km s$^{-1}$ Mpc$^{-1}$).
Similar breaks or upper cutoffs at substantially high luminosities have been found 
by Sivakoff et al. (2003) in NGC 4365 and NGC 4382, and by Jord${\rm \acute{a}}$n et al. (2004) 
in M87, M49 and NGC 4697. Moreover, in NGC 1600 Sivakoff et al. (2004) found that
the break is not needed to fit the observed XLF. The obvious disagreements between
these works cast doubts on the feasibility of using $L_{\rm b}$ as a reliable
standard candle for distance determinations.

On the other hand, it has been revealed that in early-type galaxies about 4\% of
the GCs host bright LMXBs, and about 18--70\% of the resolved
LMXBs are found to be associated with GCs with a preference on the optically brighter
and redder ones (e.g., Kundu et al. 2002; Sarazin et al. 2000, 2001, 2003). In most
cases, such a correlation with color is attributed to metallicity (Kundu et al. 2003).
These results imply that GCs are an efficient breeding ground for LMXBs,
as it is true in our own Galaxy (Katz 1975; Clark 1975). The origin of the LMXBs that
are not associated with GCs (field LMXBs) is still unclear. It has been speculated
in many published papers (e.g., Kundu et al. 2002, Maccarone et al 2003) that the
majority of them may be an ejected GC population, or have been left in the field
after GC disaggregation. Still, the possibility that they are a true field population
and were actually formed in situ cannot be ruled out at present (Maccarone et al. 2003).
The comparisons between the X-ray properties of the GC LMXBs and field LMXBs should
help answer the question. As of today, it appears that in some cases the difference
between the spectra of GC LMXBs and field LMXBs is almost undistinguishable in the
X-ray band, strongly supporting a similar origin. The only exception may be NGC 4472,
for which Maccarone et al. (2003) showed that the GC LMXBs tend to be slightly harder.

In this paper, we present a {\it Chandra} study of the X-ray point source population 
in the X-ray bright galaxy NGC 4552 (M89). NGC 4552 is located at $z=0.001134$
(Smith et al. 2000) in the Virgo cluster. It is usually classified as an elliptical
galaxy, although in some cases (e.g., Ferrari et al. 1999) it is considered a S0.
It is also a bright radio source, with a strong compact core and a relatively flat
spectrum (Filho et al. 2000; Wrobel \& Heeschen 1984). {\it HST} observations show
that from 1991 to 1996 the intensity of the central, unresolved source of this galaxy
changed by a factor of several in the near UV band, along with the appearance of
some UV/optical emission lines (Renzini et al. 1995; Cappellari et al. 1999). This
indicates that NGC 4552 harbors a mini-AGN at the center, as has been inferred in
the X-ray band by Colbert and Mushotzky (1999) who studied the nuclear source with
{\it ROSAT} and {\it ASCA}. Throughout this paper, by assuming the Hubble constant
to be
$H_0$ = 70 km s$^{-1}$ Mpc$^{-1}$
(O'Sullivan et al. 2001 and references therein) we adopt a distance to NGC 4552 of
17.1 Mpc, which is slightly larger than that derived by infrared surface brightness
fluctuation analysis ($15.4 \pm 1.0$ Mpc; Tonry et al. 2001). We quote errors at the
90\% confidence level unless mentioned otherwise.

\section{OBSERVATION AND DATA REDUCTION}\label{sec:obs}
NGC 4552 was observed on April 22--23, 2001 with the CCD 2, 3, 6, 7 and 8 of the
{\it Chandra} Advanced CCD Imaging Spectrometer (ACIS) for a total exposure of 56.8 ks.
The center of the galaxy was positioned on the ACIS S3 chip (CCD 7) with an offset of
$\simeq0.64^{\prime}$ from the nominal pointing for the S3 chip, so the entire galaxy
was covered by the S3 chip. The CCD temperature was $-120^{\circ}$C. The events were
telemetried in the Very Faint mode, and the data were collected with frame times of
3.1 s. In the analysis that follows, we used the CIAO 2.3 software to process the data
acquired from the S3 chip only. In order to use the latest calibration, we started with
the Level-1 data. We only kept events with {\it ASCA} grades 0, 2, 3, 4, and 6, and
removed bad pixels, bad columns, and columns adjacent to bad columns and node boundaries.
In order to identify occasional intervals of high background (``background flares''),
whose effects are particularly significant on the backside-illuminated S1 and S3 chips,
we extracted and examined lightcurves of the background regions on the S3 chip in
2.5--7 keV where the background flares are expected to be most visible. We found that the
detected intervals contaminated by the particle events that raise the count rate to over
20\% more than its mean value is less than 9\% of the total exposure. We have excluded
these intervals and used a net exposure time of 52.2 ks in our analysis. Also we corrected
the aspect offset of the observation 
($\delta$R.A. = $0.17^{\prime\prime}$
and 
$\delta$Dec. = $0.12^{\prime\prime}$).

We limited the spectral analysis to the $0.7 - 7$ keV energy band in order to avoid 
the effects of calibration uncertainties at lower energies and instrumental background 
at higher energies. We extracted all of the spectra in the pulse height-invariant (PI)
channels, and performed model fittings with XSPEC v11.2.0. Because there has been a
continuous degradation in the ACIS quantum efficiency (QE) since launch, we applied
ACISABS to correct the created ARF files before using them to fit the spectra.

\section{X-RAY IMAGE}\label{sec:image}
In Figure 1, we show the raw {\it Chandra} image of NGC 4552
in 0.3--10 keV, which has not been corrected for either exposure or background.
For a comparison, we overlap the DSS optical intensity contours on the X-ray image.
It can be clearly seen that the spatial distribution of the X-ray emissions is nearly
symmetric within about 9$^{\prime\prime}$ (0.7 kpc). Outside the nuclear region
the distribution of the X-rays is elongated roughly in the north-south
direction out to about 50$^{\prime\prime}$ (4.1 kpc), 
where it does not follow the profile of the optical light. In the region about 
$20^{\prime\prime}$ east of the center there is a lack of diffuse X-ray emission.
In Figure 2, we show the smoothed X-ray image with a minimum significance of 3
and a maximum significance of 5. In the figure we use a large circle to indicate
the region within the 4 effective radii (4 $R_{\rm e}$), where 
1 $R_{\rm e}=0.49^{\prime}$ 
or 2.4 kpc (de Vaucouleurs et al. 1992). The locations of the detected X-ray point
sources are also marked, which show a clear tendency to concentrate toward the
center. The details of the detection and analysis of these point sources are
presented in \S4. There is a lack of diffuse X-rays in the region about $20^{\prime\prime}$
east of the center, while the region about $36 ^{\prime\prime}$ south-west of
the center lacks bright X-ray point sources.

We find that  the {\it Chandra} X-ray position of the central point source
(Src 1; see \S4.6), the brightest one in the field as well as the peak of the
diffuse X-ray emission, coincides with the optical/IR center of the galaxy
(Monet et al. 1998; Cutri et al. 2003) within $0.5^{\prime\prime}$. This is
the X-ray counterpart of the mini-AGN identified through its UV/optical
activity (Cappellari et al. 1999). We do notice that by using the {\it ROSAT}
HRI data Colbert and Mushotzky (1999) reported an offset of $4.6^{\prime\prime}$
between the position of the compact X-ray source in the nuclear region and
the position of the optical photometric center of the galaxy. But since this
offset is much smaller than the uncertainty of the {\it ROSAT} HRI positioning,
which is typically $\sim10^{\prime\prime}$, our results are not in conflict.
Moreover, the {\it Chandra} X-ray position of the central source is also in
good agreement with the position of the central radio source (Nagar et al. 2002)
to within about $1^{\prime\prime}$.

\section{X-RAY POINT SOURCES}\label{sec:source}
\subsection{Detections}\label{sec:source_detection}
We detected X-ray point sources on the ACIS S3 image using the CIAO tool celldetect.
The default signal-to-noise threshold for source detection was set to be 3, and the
energy range used for detection was restricted to 0.3--10 keV for better statistics.
We have cross-checked the results both by using the CIAO tool wavdetect and by eye
in either the 0.3--10 keV or 0.7--7 keV images. We detected a total of 79 sources
that exceeded the detection threshold. Of these, 47 lie within the 4 $R_{\rm e}$
region. We estimate that the minimum detection for a point source is approximately
$3.13 \times 10^{-4}$ counts s$^{-1}$ (16 counts for 52.2 ks)
at $r \simeq 4~R_{\rm_e}$. Notice that this value is higher than that in Sivakoff et al.
(2003) and Sarazin et al. (2001) by about 45\% and 33\%, respectively. By reducing the
signal-to-noise threshold down to 2.7, 5 more sources can be detected within the
4 $R_{\rm e}$ region. Based on visual inspection two of them are clearly fake because
the photon distribution within the detection cell does not follow that of a point 
source. Considering the large uncertainties due to the sample incompleteness for 
very faint sources, in this paper we conservatively adopt a signal-to-noise 
threshold of 3 and focus our study on the
47 sources within the 4 $R_{\rm e}$ region. We list the properties of all 47 sources in
Table 1, where we sort them in the order of increasing projected distance {\it d} from 
the center of the galaxy. We arrange the columns as follows:
(1) source number; 
(2)-(3): right ascension and declination (J2000);
(4) projected distance {\it d} from the center of the galaxy;
(5) count rate and its error;
(6) significance of the detection;
(7) intrinsic X-ray luminosity $L_{X}$, assuming the source is located at the
distance of NGC 4552 and only subjected to the Galactic absorption
($2.56 \times 10^{20}$ cm$^{-2}$; Dickey \& Lockman, 1990; see \S4.5); 
(8)-(9): hardness ratios (see \S4.3);
and
(10) notes.

\subsection{Variability of Sources}\label{sec:source_variability}
We extracted the lightcurves for each of the 47 point sources over the duration of
the {\it Chandra} observation, excluding the intervals of strong background flares.
The extractions were made in 0.3--10 keV for better statistics, since we found
that after correcting for the background the results obtained in this energy band
are consistent with those obtained in 0.7--7 keV. We calculated the Kolmogoroff-Smirov
(KS) statistic for each of the lightcurves against the null hypothesis that
the count rate of the source plus the background is uniform over the effective exposure
time. If the count rate is temporally invariant, the cumulative fraction of the count
is expected to be a diagonal from 0 to 1. We find that two sources, Src 15 (174 counts)
and Src 28 (107 counts), have a less than 5\% probability of being invariable. In order to
examine if these results are caused by local background fluctuations, we also applied the
KS test to the lightcurves extracted from the background regions adjacent to the two
sources. In neither case can we find any significant temporal variability in the background.
Therefore, we conclude that both sources are intrinsically variable. Assuming that they
are at the distance of NGC 4552 and using the best-fit absorbed power-law spectral model
for all the resolved sources (\S4.4), we estimate that Src 15 and Src 28 have luminosities
of 
$1.17^{+0.11}_{-0.30} \times 10^{39}$ ergs s$^{-1}$ 
and 
$5.12^{+1.00}_{-1.01} \times 10^{38}$ ergs s$^{-1}$, 
respectively. In Figure 3, we present their background-corrected lightcurves that show
clear variations on $1.5-2$ hr timescales, together with the lightcurve of the central
source (Src 1), which is the brightest one in the field. In terms of the KS test, the
temporal variability of Src 1 is less significant than Src 15 and Src 28. This is likely
because K-S tests are most sensitive around the median value of the independent variable.
However, a visual examination of the lightcurve suggests that it is variable on timescales
of about 1 hr or more, with flux variations of about $50\%$. We attempted Fourier analysis
techniques to detect any potential periods on minute-hour timescales for Src 1, 15 and 28.
No statistically meaningful periodicity is found in any of the sources.

\subsection{X-Ray Hardness Ratios}\label{sec:source_hardness}
As the count rates of the resolved point sources are typically quite low directly
analyzing the spectrum of the sources by fitting models is not practical. Instead
we study the hardness ratios. Following Sarazin et al. (2000) and other authors,
we measured the background-subtracted counts for each of the 47 resolved sources
in three energy bands: soft (S), 0.3--1.0 keV; medium (M), 1.0--2.0 keV; and hard (H),
2.0--10.0 keV and calculated the hardness ratios R21 and R31 using the definitions 
R21 = (M $-$ S)/(M + S) 
and 
R31 = (H $-$ S)/(H + S), 
respectively. We list the calculated hardness ratios and their 1$\sigma$ errors,
which are very large in general, in columns 8 and 9 of Table 1, and plot R31 versus
R21 for all the sources in Figure 4. Only the typical 1$\sigma$ error bars
are illustrated in order not to complicate the figure. In the same figure, we also show
the predicted hardness ratios for an absorbed power-law model with different
column densities $N_{\rm H}$ and photon indices $\Gamma$. It can be seen in
that most of the sources are located
in a diagonal band from ($-0.217$, $-0.600$) to (+0.757, +0.837). We note that 
these colors are similar to those found in the X-ray bright elliptical galaxy NGC 720 
(Jeltema et al. 2003) and NGC 4649 (Randall et al. Irwin 2004). Three sources,
i.e., Src 17 (+0.678, +0.372), Src 31 (+0.757, +0.837), and Src 45 (+0.702, +0.436),
appear to have been absorbed by a column density larger than the Galactic value.
Since they are all located far away from the center of the galaxy
($d> 30^{\prime\prime}$),
they are probably unrelated background AGNs. However, we find that the hardness
ratios of Src 17 and Src 45 cannot be reproduced by using a simple absorbed power-law
spectral model, which is not in agreement with the spectrum of the hard X-ray cosmic
background (Mushotzky et al. 2000). The brightest source in the field (Src 1) is
located at the nucleus of the galaxy and has the X-ray color
(R21, R31) = ($-0.086$, $-0.504$).
Unlike in NGC 720 (Jeltema et al. 2003) and NGC 4649 (Randall et al. 2004), we found no
supersoft sources (SSSs) in NGC 4552. Within the 4 $R_{\rm e}$ region the total X-ray
emission (diffuse emission plus all the resolved point sources) of the galaxy has a
hardness ratio of
(R21, R31) = ($-0.453 \pm 0.011$, $-0.779 \pm 0.015$),
as compared to the
(R21, R31) = ($+0.036 \pm 0.031$, $-0.221 \pm 0.033$)
hardness ratio for the resolved sources and
(R21, R31) = ($-0.565 \pm 0.012$, $-0.896 \pm 0.018$)
for the unresolved diffuse emission from the gas plus unresolved point sources
(see also \S4.4).

\subsection{Spectral Properties}\label{sec:source_spectra}
Within the 4 $R_{\rm e}$ region, there are 12 X-ray point sources
(Src 1, 2, 6, 8, 11, 15, 24, 28, 37, 41, 43 and 46)
each having more than 100 counts. We have extracted and studied the individual
spectra of these bright sources. Among them, Src 1 is the brightest and coincides
with the mini-AGN identified in the UV/optical bands (Cappellari et al. 1999).
We will present detailed analysis of its spectrum separately in \S4.6.
 We divide the rest of sources with less than 100 counts into four
groups (Table 2). Sources in group A are the hardest ones that
appeared at the top of Figure 4. Sources in group B are those
for which the Galactic column density is required when an absorbed power-law model
is applied to describe their hardness ratios, while for sources in group C a
larger column density is needed. The rest of the sources are in group D. We have
extracted and studied the cumulative spectra of each of the four groups, as well
as that of all the resolved sources (excluding Src 1). In the model fittings, the
background spectra were extracted from the local background fields adjacent to
the regions where the source spectra were extracted. The fitting results are
summarized in Table 2. As can be seen, excluding the central
source, the cumulative spectrum of all resolved sources within the
4 $R_{\rm e}$ region can be fitted best by an absorbed power-law model with
a photon index of 
$\Gamma=1.56 \pm 0.07$
when the column density is fixed to the Galactic value. The deduced intrinsic
luminosity of all the resolved sources in 0.3--10 keV is 
$1.83 \pm 0.09 \times 10^{40}$ ergs s$^{-1}$, 
assuming that the sources are all at the distance of NGC 4552.

We also attempted to separate the emission of the unresolved point sources from
the hot diffuse gas. In order to do this we extracted the spectrum of the
total diffuse emission by excluding all resolved point sources. The background for 
this spectrum was extracted on the S3 chip as far away as possible from the galaxy.
To fit the spectrum, we used a thermal component (apec model) to represent 
the contribution of the hot plasma, and a non-thermal component (power-law model) 
to represent the emission from the unresolved point sources, under the assumption 
that the hard spectral component seen in the diffuse emission is mainly due to 
the LMXBs. Both spectral components are subjected to a common absorption $N_{\rm H}$.
Because we find that when $N_{\rm H}$ is free the obtained value is only slightly
lower than, but still consistent with the Galactic value, we simply fix $N_{\rm H}$
to the Galactic value. We fixed the power-law photon index of the hard component
to the value of 1.56 measured for the cumulative spectrum of all resolved
sources; allowing the photon index to vary did not improve the fit. The best-fit gas
temperature is found at
$kT = 0.51 \pm 0.02$ keV,
and the calculated luminosity of the total diffuse emission in 0.3--10 keV is
$L_{\rm dif} = 4.32 \pm 0.22 \times 10^{40}$ ergs s$^{-1}$,
of which about 31\%, or
$1.34 \pm 0.07 \times 10^{40}$ ergs s$^{-1}$,
can be ascribed to the unresolved point sources. If calculated in count flux, 
the contribution of the unresolved sources is about 20\%. Thus, by adding the emissions
of both resolved and unresolved point sources together, we estimate that in 0.3--10
keV the LMXBs contribute about 48\% of the galaxy's total luminosity, or about
$29\%$ of the total count flux.

Next we fit the cumulative spectra of the four groups of resolved point
sources with an absorbed power-law model and list the results in
Table 2. We find that for group A the obtained photon index
is $0.74 \pm 0.27$ if the column density is fixed to the Galactic value.
The goodness of fit is marginally acceptable at $\chi^{2}_{r}=15.7/13$.
When we allow the column density to vary, the fit can be improved significantly
($\chi^{2}_{r}=7.2/12$) in terms of the F-test which yields a best-fit photon index of
$1.83^{+0.85}_{-0.55}$, 
and a large absorption of
$N_{\rm H}=6.53^{+5.40}_{-2.10} \times 10^{21}$ cm$^{-2}$.
The spectra of group B and C can be well fit by the absorbed power-law model
with $\Gamma=1.49^{+0.18}_{-0.17}$ and $2.19^{+0.56}_{-0.32}$, respectively, 
when $N_{\rm H}$ is fixed to the Galactic value for group B and allowed to vary 
for group C. The spectrum of group D appears to be complicated and cannot be easily 
explained by using an one-component spectral model.

\subsection{X-Ray Luminosity Function}\label{sec:source_luminosities}
Based on the best-fit power-law model for the cumulative spectrum of all resolved
sources within the 4 $R_{\rm e}$ region (excluding the central source; $\Gamma=1.56$), 
we convert the observed count rates of the 46 off-center sources into unabsorbed
0.3--10 keV luminosities, assuming that the sources are all at the distance of
NGC 4552. The conversion factor is 
$5.18 \times 10^{36}$ ergs cts$^{-1}$,
and the resulting luminosities range from 
$7 \times 10^{37}$ 
to 
$1.5 \times 10^{39}$ ergs s$^{-1}$.
With these results we construct the XLF, in which the central
source is not included.

First, we fit the observed cumulative XLF that is not corrected for the effect 
of incompleteness at the faint end of the luminosity function. We use the maximum-likelihood
method and determine the 90\% confidence errors by performing Monte-Carlo
simulations. Based on deep {\it Chandra} observations of blank fields
(Mushotzky et al. 2000), we estimate that within the 4 $R_{\rm e}$ region of
NGC 4552, $^{<}_{\sim}4$ of the 47 detected point sources are expected to be
unrelated background/foreground X-ray sources. In the fittings, we perform
Monte-Carlo simulations to simulate the contribution of these background/foreground
sources accordingly. We adopt either a single or a broken power-law model to
fit the XLF. The single power-law profile is expressed as
\begin{equation}\label{eq:power-law}
N(\geq L_{38}) = N_{0} L_{38}^{-\alpha},
\end{equation}
where $L_{38}$ is the 0.3--10 keV luminosity in units of 
$10^{38}$ ergs s$^{-1}$.
The broken power-law profile can be expressed in the form of Eq.(1) at the high
and low luminosities with different slopes 
$\alpha_{\rm h}$
and
$\alpha_{\rm l}$,
respectively. In the single power-law fitting, we obtain a slope 
$\alpha = 0.95^{+0.11}_{-0.2}$. However, the fit is poor with a probability of $P=35\%$
that the data and model are drawn from the same distribution. It overestimates the data at
 high luminosities and underestimates it at low luminosities. On the
other hand, the broken power-law model can significantly improve the fit
($P=74\%$). The obtained break luminosity is  
at $L_{\rm b} = 3.7^{+1.4}_{-1.9} \times 10^{38}$ ergs s$^{-1}$, 
and the two slopes are
$\alpha_{\rm h} = 2.00^{+1.68}_{-0.71}$ 
and 
$\alpha_{\rm l} = 0.73^{+0.19}_{-0.41}$,
respectively.

Next we examine the effect of incompleteness at the faint end of the observed XLF 
by adopting a method similar to that outlined in Kim and Fabbiano (2004). With the
use of the MARX package (Wise et al. 2003) we run a series of Monte-Carlo simulations
to create fake X-ray point sources in NGC 4552 with given luminosities that cover
the observed XLF's range, and then determine how many of them can be detected with
the same technique described in \S4.1. We assume that
the radial distribution of the fake sources at a given luminosity follows the standard
r$^{1/4}$ law (de Vaucouleurs, 1948). With the obtained pick-out ratios at each given
luminosity, we are able to correct both the observed XLF and the background. The
results of the fittings for the corrected XLF are shown in Figure 5. The single power-law
model obviously gives a better fit to the corrected XLF than to the uncorrected one with
a steeper slope
$\alpha = 1.18^{+0.13}_{-0.16} $
and a probability of 48\% that the data and model are drawn from the same distribution.
The broken power-law model gives the best fit ($P=82\%$), with
$L_{\rm b} = 4.4^{+2.0}_{-1.4} \times 10^{38}$ ergs s$^{-1}$, 
$\alpha_{\rm h} = 2.28^{+1.72}_{-0.53}$,
and 
$\alpha_{\rm l} = 1.08^{+0.15}_{-0.33}$.
We notice that at the 90\% confidence level these best-fit parameters agree with those
for the uncorrected XLF. For both the corrected and uncorrected XLFs, $L_{\rm b}$ and
$\alpha_{\rm h}$ are relatively poorly determined due to the small number statistics.

We also have attempted to fit the corrected XLF with a cutoff power-law model.
Whether or not the incompleteness correction is made, we find that the model gives 
poor fit to the data and actually can be rejected at the 90\% confidence level.
These results agree with those in Kim and Fabbiano (2004), where NGC 4552 is not 
included in the sample.

\subsection{Central Source}\label{sec:source_center}
Src 1 is the brightest X-ray source in the field. It has been speculated that it might be 
a mini-AGN based on observations in the optical/UV bands (Renzini et al. 1995; Cappellari et al. 1999).
As previously noted in \S3 and \S4.2, our studies reveal that 
 it is variable on the 1 hr and larger timescales,
and its X-ray position agrees very well with the IR, optical, UV and radio
centers of the galaxy. We have extracted the {\it Chandra} ACIS spectrum of this
source and applied an absorbed power-law model to it. We find that the fit
(Table 2) is acceptable with a best-fit photon index
$\Gamma = 2.11^{+0.19}_{-0.18}$.
This confirms the result of Colbert and Mushotzky (1999), who obtained
$\Gamma = 2.24^{+0.20}_{-0.18}$
with {\it ASCA} data. Using the best-fit parameters, we estimate that in
0.3--10 keV the central source has an unabsorbed luminosity of
$3.99 \pm 0.44 \times 10^{39}$ ergs s$^{-1}$.
Based on these multi-band parameters we conclude that the central source is most likely a
low-luminosity AGN rather than a clump of LMXBs.

\subsection{Off-Center Sources with $L_{\rm X} > 10^{39}$ erg s$^{-1}$}\label{sec:source_brightest}
We detected three off-center X-ray sources (Src 15, 41 and 43) that have
luminosities larger than $10^{39}$ erg s$^{-1}$. One of them (Src 15) is located
 at about $26^{\prime\prime}$ from the center of the galaxy, and has temporal
variability on $^{>}_{\sim} 1.5$ hr timescales. The other two are at about
$90^{\prime\prime}$ from the center with no evidence for variability during
the observation. Src 41 is located in the joint {\it Chandra}-{\it HST} field
and is found to be associated with a GC (\S4.8).

We extracted the ACIS spectrum for each of the three sources and fitted them
with both an absorbed power-law model (PL) and an absorbed multicolor disk blackbody
model (DBB). For Src 41, when the absorption is fixed to the Galactic value,
the DBB model gives an acceptable fit ($\chi^{2}_{r} = 11.5/10$) with a disk
surface temperature at the inner radius of
$kT = 0.75^{+0.14}_{-0.12}$ keV.
The PL model gives a worse fit with a somewhat steep slope of 
$\Gamma = 2.11^{+0.21}_{-0.19}$.
If the absorption is left free, the goodness of the PL fit can be significantly
increased, but the resulting absorption column density is about one order of magnitude 
larger than the Galactic value.
For Src 15, both the PL model and DBB model give an acceptable fit. The best-fit
power-law photon index and inner disk surface temperature are
$\Gamma = 1.25^{+0.32}_{-0.29}$.
and
$kT = 1.86^{+1.38}_{-0.62}$ keV,
respectively, when the Galactic absorption value is used.
For Src 43, when the absorption is fixed to the 
Galactic value, the DBB model gives a relatively poor fit 
($\chi^{2}_{r}=31.8/22$) with an inner disk surface temperature of
$kT = 1.17^{+0.32}_{-0.24}$ keV,
while the PL model gives a slightly better fit ($\chi^{2}_{r}=27.1/22$) with a photon
index of
$\Gamma = 1.62^{+0.17}_{-0.21}$. 
As both models have residuals at about 1 keV we allowed the absorption
to be a free parameter and also introduced an additional soft spectral component in
the fits. However, neither approach improved the fits.

\subsection{ASSOCIATION WITH GLOBULAR CLUSTERS}\label{sec:gc}
Kundu and Whitmore (2001) studied the globular cluster system of NGC 4552
in their analysis of 28 elliptical galaxies using the high spatial resolution
images of the Wide Field Planetary Camera 2 (WFPC2) on board the
{\it HST}. For NGC 4552 the field of view of the selected {\it HST} pointing
lies entirely in the 4$R_{\rm e}$ region and covers an area of approximately
5 arcmin$^{2}$ (42\% of the 4 $R_{\rm e}$ region), which includes the nucleus and
the inner parts of the galaxy as well as a region out to about $2^{\prime}$
southeast to the center (Fig.2). Twenty five of the off-center X-ray sources and 210 GC
candidates with colors between $0.5 < V-I < 1.5$ are detected in this field jointly
covered by {\it Chandra} and {\it HST}. Hereafter, we restrict our analysis
and discussions to these 25 X-ray sources and 210 GCs unless mentioned otherwise.

There was a systematic offset of about $\simeq 2^{\prime\prime}$ 
between the {\it HST} and {\it Chandra} positions. After correcting this offset 
by using the method outlined in Maccarone et al. (2003), we find that 10 X-ray 
sources match GCs within $0.5^{\prime\prime}$. Two other X-ray sources are within 
$0.7^{\prime\prime}$ and $1.0^{\prime\prime}$ of the nearest GCs, respectively,
but they are not listed in the LMXB-GC matches in this work. Based on the spatial 
distributions of the X-ray sources and GCs, we estimate that $^{<}_{\sim}0.4$ fake
LMXB-GC matches should occur at random in the selected {\it Chandra}-{\it HST} field.
Thus, we consider all of the 10 LMXB-GC matches to be real. The 10 LMXBs in
GCs have count rates ranging from
$3.38 \times 10^{-4}$ 
to 
$5.74 \times 10^{-3}$ cts s$^{-1}$. 
They show no particular pattern of concentration in spatial distribution (Fig. 2). The
one located nearest to the center of the galaxy is at a distance of 
$\simeq 15^{\prime\prime}$. One of them (Src 41) is brighter than $10^{39}$ erg s$^{-1}$
(\S4.7).

\subsubsection{GCs Hosting Bright LMXBs and Non-LMXB GCs}\label{sec:gc_gc}
In Figure 6a and b, we compare the cumulative optical luminosity distributions of
GCs that host bright LMXBs (LMXB GCs) and those that do not (non-LMXB GCs) as a
function of their $V$-band and $I$-band magnitudes (Kundu \& Whitmore 2001),
respectively. As can clearly be seen, the detected X-ray sources tend to be
associated preferentially with the optically bright GCs. In the $V$ band, the
median value of the LMXB GC distribution is 22.0 mag, while the corresponding
value of the non-LMXB GCs is 23.1 mag. Quite similar to this, in the $I$ band
the two median values are 20.9 and 22.0 for LMXB GCs and non-LMXB GCs, respectively.
Using the Mann-Whitney rank-sum tests (Mann \& Whitney 1947), we calculate that
the probability of the fraction distributions of LMXB GCs and non-LMXB GCs
following the same distribution are only
$9 \times 10^{-4}$ and $3 \times 10^{-4}$
in the $V$ and $I$ bands, respectively.

Figure 6c shows the histogram of the $V-I$ color distribution of all 210 GCs.
The distribution is broad with a strong peak at $V-I \simeq 0.95$, and is more
extended to the redder side. There is a rather weak structure at $V-I \simeq 1.12$,
but the evidence for a bimodal distribution is not statistically significant
(Kundu \& Whitmore 2001). Such a distribution pattern implies the existence of a
population of bluer, metal-poor GCs and a more dispersive population of redder,
metal-rich GCs. This is consistent with the results found by other studies
such as Neilsen and Tsvetanov (1999). We find
that the $V-I$ color distribution of LMXB GCs tends to have two corresponding
concentrations. The number of LMXB GCs in the red population is $1.9 \pm 0.4$ times
as many LMXB GCs as in the blue population. To crosscheck this we plot the cumulative
fraction distributions of LMXB GCs and non-LMXB GCs as a function of the $V-I$ color
(Fig. 6d). We see that the two distributions are different. The median colors of the
LMXB GCs and non-LMXB GCs are $V-I=$1.12 and 1.02, respectively. Using the Mann-Whitney
rank-sum test, we find that the probability of the two distributions being the same
is not significant (0.33).

\subsubsection{GC-Associated LMXBs and Field LMXBs}\label{sec:gc_lmxb}
In Figure 7a we show both the cumulative number fraction of X-ray sources associated
with GCs and that of the field sources located in the joint {\it Chandra}-{\it HST}
field as a function of the intrinsic X-ray luminosity. We find that the median luminosity
of the distribution is 
$3.04 \times 10^{38}$ erg s$^{-1}$ 
for the GC-associated X-ray sources, and 
$2.63 \times 10^{38}$ erg s$^{-1}$
for the field sources. The probability that the two distributions are the same is 0.50
with the Mann-Whitney rank-sum tests, which means that there is no significant 
difference between them. Thus X-ray sources in GCs and those not in GCs may have 
nearly the same mean X-ray luminosities.

We study the X-ray spectral properties of GC-associated and field LMXBs,
first by comparing their X-ray hardness ratios as has been defined in 
\S4.3 and plotted in Figure 4, where 
the GC-associated sources are marked with open diamonds. However, it is hard to
find any obvious difference in the distribution pattern directly on the X-ray 
color-color diagram. So we present the distributions of the source number 
as a function of the R21 and R31 colors for the GC-associated and field LMXBs, 
respectively (Fig. 7b and c). We find that the median values of the
GC-associated sources are 
R21=$+0.09$ and R31$=+0.06$,  
which are slightly smaller than those of the field sources 
(R21=$+0.12$ and R31=$+0.10$). 
We thus speculated that the GC-associated LMXBs may be slightly softer than
their counterparts in the field.

In order to validate our speculation in a quantitative way, we fit the cumulative 
spectra of the GC-associated and field LMXBs by using an absorbed power-law 
model. The absorption column density is fixed to the Galactic value, because it 
cannot be well constrained if it is left free. When the 68\% errors are quoted, the 
obtained photon indices are
$\Gamma_{\rm GC}$ = $1.66 \pm 0.08$
for the GC-associated sources and
$\Gamma_{\rm Field}$ = $1.42 \pm 0.08$
for the field sources, which supports that the GC-associated X-ray sources are softer.
At the 90\% confidence level, the model gives 
$\Gamma_{\rm GC}$ = $1.66 \pm 0.12$
and 
$\Gamma_{\rm Field}$ = $1.42^{+0.14}_{-0.13}$, which overlap only marginally.

None of the GC-associated sources show significant temporal variability in X-rays.

\section{DISCUSSION}\label{sec:discussion} 
\subsection{The Break on the XLF}\label{sec:dis_xlf}
The break on the XLF of point sources in early-type galaxies has been detected at
$\simeq 3-5 \times 10^{38}$ ergs s$^{-1}$
in NGC 4697 (Sarazin et al. 2000), NGC 4472 (Kundu et al. 2002), NGC 1553 (Blanton et al. 2001),
and a sample of 14 E/S0 galaxies (including NGC 4697; Kim and Fabbiano 2004).
However, in NGC 720 (Jeltema et al. 2003) it is found at a much higher luminosity of
$L_{\rm b} = 1.07^{+0.1}_{-0.1} \times 10^{39}$ ergs s$^{-1}$
for $H_0$ = 70 km s$^{-1}$ Mpc$^{-1}$.
Recent analyses of NGC 4365, NGC 4382 (Sivakoff et al. 2003), M49, M87, and NGC 4697
(Jord{\rm $\acute{a}$}n et al. 2004) suggest an upper cutoff at
$1-2 \times 10^{39}$ ergs s$^{-1}$,
rather than a break. In NGC 1600, Sivakoff et al. (2004) showed that a single
power-law profile is sufficient to describe the observed XLF. In NGC 4552,
we argue that a break at about
$4.4 \times 10^{38}$ ergs s$^{-1}$
is necessary to fit the XLF corrected for the sample incompleteness of faint sources
(\S4.5). For luminosities below
$ 5 \times 10^{38}$ erg s$^{-1}$,
the profile of the observed XLF is consistent with that expected for ultracompact binaries
(Bildsten \& Deloye 2004).

Why are the conclusions on the XLF profile so dispersed from case to case? Is there an
intrinsic universal break luminosity? Since the typical number of the detected X-ray
sources is only 50-150 per galaxy we speculate that even if there is an universal
break, the small number statistics will preclude us from measuring it. To investigate
this possibility, we carry out direct Monte-Carlo simulations to create a series of fake
XLFs whose profiles are intrinsically determined by a broken power-law profile with the
typical parameters 
$\alpha_{\rm l} = 1.0$,
$\alpha_{\rm h} = 2.0$
and 
$L_{\rm b} = 4.0 \times 10^{38}$ ergs s$^{-1}$.
For each simulated XLF, we consider 100 sources. By fitting the resulting XLFs with 
the broken power-law model, we find that at the 90\% confidence level the obtained
$L_{\rm b}$ ranges from 
$2.8 \times 10^{38}$ ergs s$^{-1}$
to
$1.2 \times 10^{39}$ ergs s$^{-1}$. 
Here the lower limit of $L_{\rm b}$ agrees well with the break luminosity found 
in galaxies such as NGC 4697, and the upper limit is consistent with the positions
of the cutoffs found in NGC 720, NGC 4365, NGC 4382 and others. Such a large dispersion
makes determination of the universal break, if any, observationally impossible.

On the theoretical aspect, an universal break on the XLF is not necessary to signify the
transition from neutron stars to black holes. The ultra compact binaries with He or C/O
donors should have doubled Eddington luminosity comparing to their counterparts with H
donors, and such systems are expected to exist in the dense GC environment in
early-type galaxies (Bildsten \& Deloye 2004). Also, LMXBs may emit above the Eddington
limit if the emission is not isotropic.

\subsection{Low Mass X-ray Binaries and Their Associations with Globular Clusters}\label{sec:dis_gc}
By analyzing the {\it Chandra} and {\it HST} data, we find that the fraction of the
GCs hosting bright LMXBs in NGC 4552 (4.8\%) is similar to those found in 
NGC 1399 (3.8\%; Angelini et al. 2001), 
NGC 4472 (4\%; Kundu et al. 2002),
NGC 1553 (2.9\%), NGC 4365 (5.5\%), NGC 4649 (4.9\%) and 
NGC 4697 (2.7\%; Sarazin et al. 2003). 
Moreover, the fraction of LMXBs associated with GCs is 40\% in NGC 4552,
which is similar to those of the X-ray bright ellipticals
NGC 4472 (40\%; Kundu et al. 2002) and
NGC 4649 (47\%), and the X-ray faint ellipticals NGC 4365 (49\%) and
NGC 4697 (44\%; Sarazin et al. 2003),
which is consistent with the argument of White et al. (2002) that at present GCs are
the dominant sites for LMXB formation in early-type galaxies. The fraction of
GC-associated LMXBs in NGC 4552 is higher than those of the S0 galaxies
NGC 1553 (18\%; Sarazin et al. 2003) and
NGC 1332 (30\%; Humphrey et al. 2004),
but smaller than that of the cD galaxy
NGC 1399 (70\%; Angelini et al. 2001).
Considering that in typical spiral galaxies such as our Galaxy and M31 (e.g., Supper et al. 1997)
the fraction is only about 10\%, these observations are generally in agreement with
previous suggestions  (e.g. Sarazin et al. 2003)  that the fraction of X-ray sources
residing in GCs may increase along the Hubble sequence from spiral bulges to S0, E,
and then cD galaxies. However, it is important to note that the LMXB-GC connection
has only been studied in small central regions observed by the HST in each of the
early type galaxies mentioned here. This apparent variation with Hubble type may be
amplified by any spatial variation in the rate of GC-LMXB associations. We note that
considering only the LMXBs in the inner few kpc of the Milky Way to the comparable
HST based analyses of the more distant early type galaxies would suggest similar
GC-LMXB association rates in early and late type galaxies.

We calculate that in NGC 4552 the probability of having bright X-ray sources in GCs is 
$1.47 \times 10^{-7}$ LMXB per $L_{\odot,I}$,
which agrees very well with the values obtained by Kundu et al. (2003) and Sarazin et al. (2003).
Bildsten \& Deloye (2004) argue that this can be explained with ultracompact binaries that
have a birthrate of one new mass transferring binary every $2 \times 10^6$ yr per $10^{7}~M_{\odot}$
of GCs. We find that GCs hosting bright LMXBs are typically 1--2 magnitudes brighter than those
with no detected LMXBs in the $V$ and $I$ bands. In fact at the level of significance of $10^{-3}$
we reject the hypothesis that the luminosities of LMXB GCs and non-LMXB GCs are drawn from the
same distribution. We also find that in the red, metal-rich GCs there are about $1.9 \pm 0.4$ times
as many LMXBs as there are in the blue, metal-poor ones. These results are consistent with,
or quite similar to those found in NGC 4472 (Kundu et al. 2002), M87 (Jord{\rm $\acute{a}$}n et al. 2004)
and other S0/E galaxies (Sarazin et al. 2003), indicating that the high GC formation
efficiency is largely attributed to the metallicity, rather than the age of the old
stellar systems (Kundu et al. 2003). In NGC 4472 and M87, the number of redder GCs that
host a LMXB is about 3 times more than their bluer counterparts. The relatively lower
overabundance of LMXB ($\simeq 2$) in redder GCs in NGC 4552 may be partly due to the
fact that the redder GC population is less prominent in this galaxy
(Kundu \& Whitmore 2001).

A comparison of the X-ray properties of the GC LMXBs and field populations of LMXBs may also
help us understand the origin of the field LMXBs. In NGC 4552, we find no obvious difference
between the luminosity distributions of GC and field LMXBs. By examining both the
hardness ratios and the cumulative X-ray spectra of the GC and field sources, we find
that the LMXBs in GCs are softer than the field LMXBs at the 68\% confidence level. Neither
the GC LMXB sample nor field LMXBs has particularly bright soft or hard
sources that may dominate the counts in the combined spectrum, which can bias the analysis
significantly. Thus, the spectral difference appears to be physically real.
This may imply that in NGC 4552 the mean metallicity of the GC-associated LMXBs
is higher than that of the field LMXBs (Maccarone et al. 2004), as opposed to NGC 4472,
where the LMXBs in GCs tend to be slightly harder (Maccarone et al. 2003).

\subsection{Brightest Off-Center X-Ray Sources}\label{sec:dis_brightest}
Typically, 1--4 X-ray sources that have a 0.3--10 keV luminosity larger than 
$10^{39}$ erg s$^{-1}$
are detected in each early-type galaxy studied to date (e.g., Sarazin et al. 2001; Blanton et al. 2001;
Angelini et al. 2001; Kim \& Fabbiano 2003; Humphrey \& Buote 2004). In a few cases,
such as NGC 720 (Jeltema et al. 2003) and NGC 1600 (Sivakoff et al. 2004), an even larger
number has been reported. The nature of these very bright sources is not clear. Some
of these studies suggest that they are ultra-luminous X-ray sources (ULXs) that host an
intermediate mass black hole (IMBH). However, in a recent study of nearby galaxies Irwin et al. (2004)
showed that in early type galaxies the sources brighter than
$2 \times 10^{39}$ erg s$^{-1}$
are most probably unassociated with the galaxy, while the sources with luminosities of
$1-2 \times 10^{39}$ erg s$^{-1}$
can be explained by accretion onto 10-20 $M_{\odot}$ stellar mass black holes (SMBHs).
Also in NGC 720, when a more conservative distance to the galaxy is adopted the number of
the very bright sources associated with the galaxy is not statistically
significant.

The three brightest off-center X-ray sources detected within the 4 $R_{\rm e}$ region of
NGC 4552 have isotropic 0.3--10 keV luminosities of  
$1.18 \times 10^{39}$ ergs s$^{-1}$ (Src 15),
$1.15 \times 10^{39}$ ergs s$^{-1}$ (Src 41) 
and 
$1.54 \times 10^{39}$ ergs s$^{-1}$ (Src 43), 
respectively. One of them, Src 41, is in the joint {\it Chandra}-{\it HST} field and
is found to be associated with a globular cluster. Similar very bright X-ray source-GC
matches also have been seen in other early-type galaxies by Angelini et al. (2001; NGC 1399)
and Jeltema et al. (2003; NGC 720). We speculate that these very bright GC-associated
sources may be powered by the accretion onto a black hole that is at the center of the
host GC. Actually the X-ray spectrum of Src 41 in NGC 4552 can be better fitted with a
single multiple blackbody disk model than a power-law model (assuming Galactic
absorption). The best-fit inner disk temperature  
($0.75^{+0.14}_{-0.12}$ keV)
is higher than that of  
NGC X-1 ($kT_{\rm in} \simeq 0.1$ keV; Colbert \& Mushotzky 1999), 
M81 X-9 ($kT_{\rm in} \simeq 0.2$ keV; Miller et al. 2004)
and others, inferring that the accreting source in Src 41 should have a relatively low
mass, which is estimated to be about 
23 $M_{\odot}$ 
for a Schwarzschild black hole, or
15--135 $M_{\odot}$
for a Kerr black hole, depending on the black hole spin and the sense of the disk
rotation. This ambiguous result makes it difficult to distinguish between the IMBH and
SMBH natures for Src 41. Although it is also possible that Src 41 is a background AGN,
the probability for such an AGN-GC match is very low.

For Src 43, when the absorption is fixed to the Galactic value, the multiple blackbody
disk model gives an unacceptable fit to its X-ray spectrum. A better and marginally
acceptable fit can be obtained with the power-law model with a photon index of
$1.62^{+0.17}_{-0.21}$. 
So this source is likely to be a neutron star binary system with beamed emissions or
a SMBH binary system. For Src 15, both the power-law model and multiple blackbody model
give acceptable fits. The best-fit inner disk surface temperature 
($1.86^{+1.38}_{-0.62}$ keV)
is consistent with that of a neutron star or a SMBH system. Still, we cannot exclude the
possibility that these two source are background AGNs.

\section{SUMMARY}\label{sec:summary}
By analyzing {\it Chandra} ACIS data we have detected 47 X-ray point sources
within the inner 4 $R_{\rm e}$ region of the early-type galaxy NGC 4552. Most of
the sources are inferred to be LMXBs. The position of the brightest point
source is consistent with that of the IR, optical, UV and radio centers of
the galaxy. In the X-ray band, the central source shows a relatively steep
power-law spectrum and temporal variability on $^{>}_{\sim}$ 1 hr timescales.
These results confirm the early speculation that a low-luminosity AGN resides in
the center of this galaxy
(Renzini et al. 1995; Cappellari et al. 1999). The derived 0.3--10 keV luminosities
of the 46 off-center sources range from
$7 \times 10^{37}$ 
to
$1.5 \times 10^{39}$ erg s$^{-1}$.
Three sources have isotropic 0.3--10 keV luminosities larger than
$10^{39}$ erg s$^{-1}$.
One of them (Src 41) is in the joint {\it Chandra}-{\it HST} field and is associated
with a globular cluster, and another (Src 15) shows temporal variations on
$^{>}_{\sim} 1.5$ hr timescales. By studying their ACIS spectra we find that Src 41 may
be a black hole system with a mass of 15--135 $M_{\odot}$, while the other two sources
should have lower masses if they are associated with the galaxy and not background AGN.

We find that after correcting for the incompleteness at the low luminosity end
the observed cumulative XLF can be best fit by a broken power-law model with a
break at
$L_{\rm b} = 4.4^{+2.0}_{-1.4} \times 10^{38}$ ergs s$^{-1}$, 
while the single power-law model and the cutoff power-law model give worse, and
unacceptable, fits. The position of the break is consistent with that found
by Kim and Fabbiano (2004) in a sample of 14 E/S0 galaxies. By performing
Monte-Carlo simulations we argue that even if there is an universal break, it
is not a reliable distance indicator due to small number
statistics.

In an area  jointly covered by both the {\it Chandra}
ACIS and {\it HST} WFPC2, we detected 25 off-center X-ray point sources and 210 GCs,
including 10 LMXB-GC matches. We find that the fraction of the GCs hosting bright LMXBs
(4.8\%) and the fraction of LMXBs associated with GCs (40\%) are both in good
agreement with those in other early-type galaxies (e.g., Kundu et al. 2002;
Sarazin et al. 2003). As in NGC 4472 (Kundu et al. 2002)
and M87 (Jord{\rm $\acute{a}$}n et al. 2004) and other early type galaxies,
in NGC 4552 the GCs hosting bright LMXBs
are typically 1--2 magnitudes brighter than the GCs with no detected LMXBs in the $V$
and $I$ bands. Moreover, there are about $1.9 \pm 0.4$ times
as many LMXBs  in the red, metal-rich GCs  as there are in the blue, metal-poor ones.
This supports the idea
that the high GC formation efficiency is largely attributed to the metallicity
in old stellar systems (Kundu et al. 2003). We find no obvious difference
between the X-ray luminosity distributions of GC LMXBs and field LMXBs. The cumulative
spectrum of the LMXBs in GCs tend to be softer than that of the field LMXBs, which
differs from result of Maccarone et al. (2003) who showed that in NGC 4472 the
LMXBs in GCs tend to be slightly harder. This may indicate that in this galaxy the mean metallicity
of the GC-associated LMXBs is higher than that of the field LMXBs (Maccarone et al. 2004).

\acknowledgments
This work was supported by the National Science Foundation of China
(Grant No. 10273009 and 10233040), and by the Ministry of Science and
Technology of China, under Grant No. NKBRSF G19990754. AK thanks NASA
for support via LTSA grant NAG5-12975.

\newpage

\clearpage
\begin{deluxetable}{lccrrrcrrr}
\tabletypesize{\scriptsize}
\tablecaption{X-Ray Properties of the Point Sources \label{tbl:sources}}
\tablewidth{0pt}
\tablecolumns{10}
\tablehead{
\colhead{Src} &
\colhead{R.A.} &
\colhead{Dec.} &
\colhead{d\tablenotemark{a} \,} &
\colhead{Count Rate\tablenotemark{b} \,} &
\colhead{S/N} &
\colhead{Luminosity\tablenotemark{c} \,} &
\colhead{R21\tablenotemark{d} \,} &
\colhead{R31\tablenotemark{e} \,} &
\colhead{Notes\tablenotemark{f} \,} \\
\colhead{No.} &
\colhead{(J2000)} &
\colhead{(J2000)} &
\colhead{($^{\prime\prime}$)} &
\colhead{(10$^{-4}$ s$^{-1}$)} &
\colhead{($\sigma$)} &
\colhead{(10$^{37}$ ergs s$^{-1}$)}
}
\startdata
1 &12:35:39.82  &+12:33:23.0 &0.0  &180.9$\pm$6.6  &19.6 &338.5  &$-0.09\pm0.06$ &$-0.50\pm0.05$ &AGN, V\\
2 &12:35:39.92  &+12:33:18.0 &5.2  & 16.4$\pm$2.4  &6.2  &36.2   &$+0.12\pm0.25$ &$-0.20\pm0.26$ &\\
3 &12:35:39.69  &+12:33:18.1 &5.2  &  9.2$\pm$2.1  &3.5  &28.0   &$+0.12\pm0.49$ &$+0.23\pm0.42$ &B\\
4 &12:35:40.30  &+12:33:10.7 &14.1 &  9.2$\pm$1.7  &4.6  &27.4   &$-0.16\pm0.28$ &$-0.45\pm0.26$ &B\\
5 &12:35:40.53  &+12:33:33.3 &14.6 &  7.5$\pm$1.7  &4.3  &21.9   &$+0.15\pm0.58$ &$+0.44\pm0.40$ &B, GC\\
6 &12:35:40.62  &+12:33:12.2 &15.9 & 19.0$\pm$2.0  &8.1  &56.6   &$+0.00\pm0.19$ &$+0.03\pm0.19$ &\\
7 &12:35:40.90  &+12:33:17.1 &16.8 & 15.6$\pm$1.8  &6.8  &34.5   &$+0.31\pm0.19$ &$-0.04\pm0.24$ &C\\
8 &12:35:39.37  &+12:33:41.9 &20.0 & 24.2$\pm$2.3  &9.0  &58.7   &$+0.00\pm0.15$ &$-0.23\pm0.16$ &\\
9 &12:35:40.63  &+12:33:40.5 &21.2 &  4.5$\pm$1.0  &3.9  &13.3   &$+0.13\pm0.45$ &$+0.17\pm0.43$ &B\\
10 &12:35:38.32 &+12:33:18.1 &22.6 &  7.2$\pm$1.3  &4.9  &14.4   &$-0.36\pm0.26$ &$-0.33\pm0.26$ &D\\
11 &12:35:39.69 &+12:33:00.4 &22.6 & 19.6$\pm$2.0  &7.9  &61.6   &$+0.24\pm0.16$ &$-0.07\pm0.20$ &GC\\
12 &12:35:40.76 &+12:33:02.8 &24.4 &  5.5$\pm$1.1  &4.1  &16.1   &$+0.21\pm0.34$ &$+0.12\pm0.36$ &B\\
13 &12:35:39.93 &+12:33:48.1 &25.2 &  4.4$\pm$1.2  &3.2  &8.3    &$-0.54\pm0.43$ &$-0.12\pm0.37$ &D\\
14 &12:35:38.06 &+12:33:25.6 &25.9 & 12.4$\pm$1.6  &6.2  &24.6   &$-0.25\pm0.20$ &$-0.36\pm0.20$ &D\\
15 &12:35:38.07 &+12:33:28.3 &26.2 & 31.7$\pm$2.5  &10.2 &117.7  &$+0.16\pm0.13$ &$+0.03\pm0.14$ &V\\
16 &12:35:38.00 &+12:33:37.4 &30.3 &  5.0$\pm$1.0  &3.7  &15.0   &$-0.13\pm0.32$ &$-0.55\pm0.30$ &B\\
17 &12:35:38.24 &+12:33:02.8 &30.7 &  6.4$\pm$1.2  &4.4  &26.4   &$+0.68\pm0.25$ &$+0.37\pm0.46$ &A\\
18 &12:35:40.50 &+12:32:50.5 &33.9 & 13.8$\pm$1.7  &6.6  &31.3   &$+0.07\pm0.20$ &$-0.30\pm0.22$ &C, GC\\
19 &12:35:40.27 &+12:32:49.1 &34.4 &  5.2$\pm$1.1  &4.0  &10.3   &$-0.43\pm0.34$ &$-0.29\pm0.31$ &D\\
20 &12:35:40.47 &+12:33:57.1 &35.4 &  3.4$\pm$0.9  &3.0  &7.6    &$-0.09\pm0.38$ &$-0.81\pm0.34$ &C\\
21 &12:35:40.16 &+12:33:58.1 &35.5 & 11.9$\pm$1.5  &6.3  &34.9   &$+0.25\pm0.23$ &$+0.22\pm0.23$ &B\\
22 &12:35:37.33 &+12:33:22.2 &36.5 &  4.2$\pm$0.9  &3.5  &12.0   &$+0.08\pm0.35$ &$-0.18\pm0.41$ &B\\
23 &12:35:40.99 &+12:32:48.4 &38.6 &  3.4$\pm$0.9  &3.0  &10.3   &$-0.22\pm0.37$ &$-0.60\pm0.36$ &B, GC\\
24 &12:35:41.37 &+12:32:51.1 &39.0 & 19.4$\pm$2.0  &8.0  &73.2   &$+0.19\pm0.17$ &$+0.04\pm0.19$ &\\
25 &12:35:42.17 &+12:33:00.5 &41.0 &  3.7$\pm$0.9  &3.2  &10.8   &$-0.04\pm0.37$ &$-0.33\pm0.38$ &B, GC\\
26 &12:35:41.61 &+12:32:51.1 &41.2 &  5.4$\pm$1.0  &3.8  &12.0   &$-0.03\pm0.29$ &$-0.55\pm0.29$ &C\\
27 &12:35:41.30 &+12:33:59.6 &42.5 & 10.0$\pm$1.4  &5.6  &29.4   &$+0.11\pm0.25$ &$+0.03\pm0.26$ &B, GC\\
28 &12:35:36.90 &+12:33:08.0 &45.3 & 19.4$\pm$2.0  &7.5  &52.6   &$-0.04\pm0.16$ &$-0.21\pm0.17$ &V\\
29 &12:35:36.91 &+12:33:05.6 &46.0 &  4.1$\pm$0.9  &3.3  &7.6    &$-0.40\pm0.35$ &$-0.06\pm0.35$ &D\\
30 &12:35:42.39 &+12:32:54.1 &47.3 & 11.5$\pm$1.5  &5.9  &25.5   &$+0.32\pm0.20$ &$+0.03\pm0.27$ &C\\
31 &12:35:43.40 &+12:33:10.0 &53.9 & 11.2$\pm$1.5  &5.4  &47.8   &$+0.76\pm0.17$ &$+0.84\pm0.11$ &A\\
32 &12:35:41.76 &+12:32:35.7 &55.1 &  4.7$\pm$1.0  &3.4  &9.1    &$-0.18\pm0.33$ &$-0.32\pm0.34$ &D\\
33 &12:35:36.03 &+12:33:33.7 &56.6 & 13.0$\pm$1.6  &6.4  &37.4   &$-0.01\pm0.21$ &$-0.11\pm0.21$ &B\\
34 &12:35:35.25 &+12:33:07.4 &68.8 &  6.1$\pm$1.1  &4.1  &13.3   &$+0.11\pm0.26$ &$-0.70\pm0.27$ &C\\
35 &12:35:36.11 &+12:32:38.9 &70.0 &  3.7$\pm$0.9  &3.0  &10.7   &$-0.00\pm0.40$ &$-0.05\pm0.40$ &B\\
36 &12:35:39.69 &+12:34:33.1 &70.2 &  6.0$\pm$1.1  &4.3  &11.3   &$-0.10\pm0.32$ &$-0.03\pm0.32$ &D\\
37 &12:35:44.83 &+12:33:41.0 &75.6 & 24.2$\pm$2.2  &8.7  &71.9   &$+0.02\pm0.15$ &$-0.10\pm0.16$ &GC\\
38 &12:35:38.18 &+12:32:03.3 &83.2 &  5.2$\pm$1.0  &4.1  &15.1   &$+0.16\pm0.33$ &$+0.08\pm0.35$ &B, GC\\
39 &12:35:45.69 &+12:33:22.4 &85.9 & 16.2$\pm$1.8  &7.9  &47.1   &$-0.04\pm0.18$ &$-0.22\pm0.19$ &B, GC\\
40 &12:35:38.49 &+12:31:56.5 &88.6 &  9.9$\pm$1.4  &6.0  &18.2   &$-0.08\pm0.23$ &$-0.19\pm0.25$ &D\\
41 &12:35:45.79 &+12:33:02.9 &89.7 & 57.4$\pm$3.4  &15.8 &114.7  &$+0.15\pm0.09$ &$-0.38\pm0.11$ &GC\\
42 &12:35:33.69 &+12:33:37.5 &91.0 &  3.6$\pm$0.9  &3.2  &10.5   &$+0.15\pm0.40$ &$-0.03\pm0.45$ &B\\
43 &12:35:41.23 &+12:34:51.7 &91.1 & 58.6$\pm$3.4  &15.8 &154.0  &$+0.15\pm0.09$ &$-0.12\pm0.11$ &\\
44 &12:35:43.19 &+12:34:50.0 &100.1 & 8.7$\pm$1.3  &5.2  &20.6   &$+0.18\pm0.23$ &$-0.26\pm0.31$ &C\\
45 &12:35:39.23 &+12:31:42.6 &100.7 & 6.5$\pm$1.1  &3.8  &27.1   &$+0.70\pm0.20$ &$+0.44\pm0.36$ &A\\
46 &12:35:41.53 &+12:31:39.6 &106.3 &27.5$\pm$2.4  &8.8  &63.7   &$+0.03\pm0.13$ &$-0.39\pm0.14$ &\\
47 &12:35:46.63 &+12:32:40.1 &108.5 & 3.1$\pm$0.8  &3.2  &7.0    &$+0.16\pm0.38$ &$-0.50\pm0.50$ &C\\
\enddata
\tablenotetext{a}{Projected distances measured from the center of the galaxy.}
\tablenotetext{b}{Background-subtracted count rates extracted in 0.3--10 keV and their 1 $\sigma$ errors.}
\tablenotetext{c}{Unabsorbed 0.3--10 keV luminosities calculated by using the
best-fit spectral model, assuming that the sources are located at the distance of NGC 4552.}
\tablenotetext{d-e}{Hardness ratios R21 = (M $-$ S)/(M $+$ S) and R31 = (H $-$ S)/(H $+$ S),
calculated by using the background-subtracted counts in 0.3--1.0 keV (S), 1.0--2.0 keV (M)
and 2.0--10 keV (H).}
\tablenotetext{f}{AGN -- Active galactic nuclei;
A, B, C, and D -- Sources in groups A, B, C, and D, respectively; 
GC -- Sources associated with globular clusters;
V -- Sources showing significant temporal variability.}
\end{deluxetable}

\clearpage
\begin{deluxetable}{lccccccccr}
\tabletypesize{\tiny}
\tablecaption{Results of the Spectral Fittings \label{tbl:spectra}}
\tablewidth{0pt}
\tablehead{
& & &
\multicolumn{2}{c}{Non-thermal Component} & &
\multicolumn{3}{c}{Thermal Component} &\\
\cline{4-5}
\cline{7-9}
\colhead{Source} &
\colhead{Model\tablenotemark{a} \,} &
\colhead{$N_{\rm H}$} &
\colhead{$\Gamma$ or $kT_{\rm in}$\tablenotemark{b} \,} &
\colhead{$F_{\rm X}$\tablenotemark{c} \,} & &
\colhead{$kT$\tablenotemark{d} \,} &
\colhead{$Z$\tablenotemark{e} \,} &
\colhead{$F_{\rm X}$\tablenotemark{c} \,} &\\
& &
\colhead{($10^{20}$ cm$^{-2}$)} &
\colhead{(keV)} & & &
\colhead{(keV)} &
\colhead{(solar)} & &
\colhead{$\chi^2$/d.o.f}
}
\startdata
\hline
\multicolumn{10}{c}{(1) Resolved and Unresolved Emissions}\\
\hline
Resolved Sources\tablenotemark{f} \, &PL &2.56/fixed             &$1.56^{+0.07}_{-0.07}$ &5.26 & & & & &106.1/97\\
Diffuse\tablenotemark{g} \, &PL+APEC &2.56/fixed &1.56/fixed &3.85 & &0.51$^{+0.02}_{-0.02}$ &$0.82^{+3.41}_{-0.10}$ &8.55 &106.4/65\\
                            &PL+APEC &$< 1.16$   &1.56/fixed &3.82 & &0.52$^{+0.03}_{-0.02}$ &$0.59^{+1.24}_{-0.26}$ &7.95 &100.1/64\\
Total &PL+APEC &2.56/fixed &1.56/fixed &10.76 & &$0.55^{+0.01}_{-0.05}$ &0.82/fixed &8.70 &160.3/136\\
      &PL+APEC &$< 0.71$   &1.56/fixed &10.85 & &$0.56^{+0.01}_{-0.01}$ &0.82/fixed &7.84 &150.4/135\\
\hline\hline
\multicolumn{10}{c}{(2) Resolved Point Sources in 4 Groups}\\
\hline
Group A &PL       &$65.3^{+54.0}_{-21.0}$ &$1.83^{+0.85}_{-0.55}$ &0.26 & & & & &7.2/12\\
Group B &PL       &2.56/fixed             &$1.49^{+0.18}_{-0.17}$ &1.06 & & & & &25.1/22\\
Group C &PL       &$19.7^{+8.29}_{-13.7}$ &$2.19^{+0.56}_{-0.32}$ &0.44 & & & & &13.5/12\\
Group D &PL       &$< 1.00$               &$1.62^{+0.38}_{-0.31}$ &0.35 & & & & &41.7/21\\
        &PL+APEC  &$< 0.81$               &$0.96^{+0.25}_{-0.23}$ &0.41 & &$0.58^{+0.18}_{-0.20}$ &0.82/fixed &0.05 &27.6/19\\
\hline\hline
\multicolumn{10}{c}{(3) Individual Sources\tablenotemark{h} \,}\\
\hline
\multicolumn{10}{c}{(3.1) Central Source}\\
\hline
Src 1 &PL &2.56/fixed &$2.11^{+0.19}_{-0.18}$ &1.15 & & & & &35.2/47\\
\hline
\multicolumn{10}{c}{(3.2) Sources Showing Significant Temporal Variabilities and/or Brighter than $10^{39}$ erg s$^{-1}$\tablenotemark{i}}\\
\hline
Src 15 (V)   &PL  &2.56/fixed &$1.25^{+0.32}_{-0.29}$ &0.34 & & & & &14.5/22\\
              &DBB &2.56/fixed &$1.86^{+1.38}_{-0.62}$ &0.26 & & & & &15.3/22\\
Src 28 (V)   &PL  &2.56/fixed &$1.56^{+0.51}_{-0.45}$ &0.15 & & & & &10.4/18\\
Src 41       &PL  &2.56/fixed             &$2.11^{+0.21}_{-0.19}$ &0.43 & & & & &18.4/10\\
              &PL  &$37.0^{+18.7}_{-15.7}$ &$2.94^{+0.67}_{-0.36}$ &0.29 & & & & &9.3/9\\
              &DBB &2.56/fixed             &$0.75^{+0.14}_{-0.12}$ &0.29 & & & & &11.5/10\\
Src 43       &PL  &2.56/fixed &$1.62^{+0.17}_{-0.21}$ &0.52 & & & & &27.1/22\\
              &DBB &2.56/fixed &$1.17^{+0.32}_{-0.24}$ &0.36 & & & & &31.8/22\\
\hline
\multicolumn{10}{c}{(3.3) Other Bright Sources}\\
\hline
Src 2  &PL &2.56/fixed &$1.85^{+0.59}_{-0.49}$ &0.15 & & & & &3.0/8\\
Src 6  &PL &2.56/fixed &$1.47^{+0.52}_{-0.44}$ &0.16 & & & & &10.9/13\\
Src 8  &PL &2.56/fixed &$1.72^{+0.47}_{-0.41}$ &0.20 & & & & &6.76/10\\
Src 11 &PL &2.56/fixed &$1.44^{+0.40}_{-0.36}$ &0.19 & & & & &6.5/12\\
Src 24 &PL &2.56/fixed &$1.23^{+0.38}_{-0.34}$ &0.21 & & & & &9.6/13\\
Src 37 &PL &2.56/fixed &$1.47^{+0.40}_{-0.35}$ &0.22 & & & & &11.5/19\\
Src 46 &PL &2.56/fixed &$1.76^{+0.46}_{-0.40}$ &0.18 & & & & &13.3/18\\
\enddata

\tablenotetext{a}{Models used in the spectral fittings:
PL = the power-law model;
APEC = the thermal plasma model;
and
DBB = the multiple color disk model.}
\tablenotetext{b}{Photon indices $\Gamma$ of the power-law model, or surface temperatures
at the inner radius of the accretion disk $kT_{\rm in}$ (in keV) of the DBB model.}
\tablenotetext{c}{Unabsorbed 0.3--10 keV fluxes in $10^{-13}$ ergs cm$^{-2}$ s$^{-1}$.}
\tablenotetext{d}{Gas temperatures of the APEC model.}
\tablenotetext{e}{Metal abundances of the APEC model in the solar unit.}
\tablenotetext{f}{All the resolved point sources excluding the central one.}
\tablenotetext{g}{Diffuse emission excluding all the resolved point sources.}
\tablenotetext{h}{Sources each having more than 100 counts.}
\tablenotetext{i}{Variable sources are marked with a V.}
\end{deluxetable}

\clearpage

\begin{figure}
\epsscale{0.6}
\begin{center}
\includegraphics[width=7.5cm,angle=0]{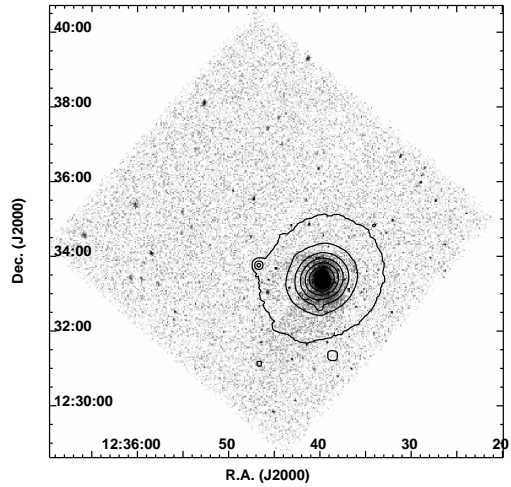}
\end{center}
\figcaption{
The raw {\it Chandra} S3 image of NGC 4552 in 0.3--10 keV, which has not been
corrected for either exposure or background. The DSS optical intensity contours
are overlaid on a linear scale. Coordinates are in J2000.
\label{fig1}}
\end{figure}

\begin{figure}
\epsscale{0.6}
\begin{center}
\includegraphics[width=7.5cm,angle=0]{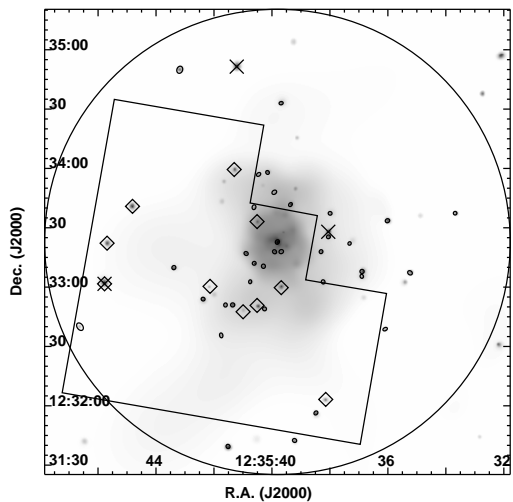}
\end{center}
\figcaption{
The X-ray image of NGC 4552 smoothed with a minimum significance of 3 and a maximum
significance of 5. The marks are: Polygon --- the {\it HST} WFPC2 field of view;
large circle --- the 4 effective radii (1 $R_{\rm e} = 0.49^{\prime}$ or 2.4 kpc;
de Vaucouleurs et al. 1992); cross: X-ray sources brighter than $10^{39}$ erg s$^{-1}$;
diamond: X-ray sources associated with GCs; small circles: other detected X-ray point sources.
All point sources were detected by using a signal-to-noise threshold of 3 for celldetect, and
coordinates are in J2000.
\label{fig2}}
\end{figure}

\begin{figure}
\epsscale{0.6}
\begin{center}
\includegraphics[width=6.0cm,angle=270]{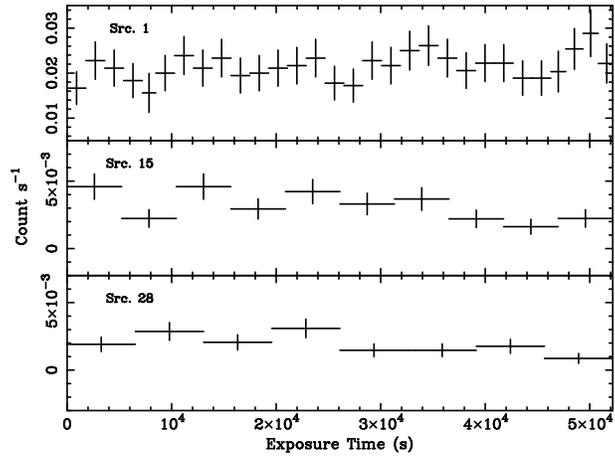}
\end{center}
\figcaption{
Background-corrected 0.3--10 keV lightcurves of the sources that show significant
temporal variability.
\label{fig3}}
\end{figure}

\begin{figure}
\epsscale{0.6}
\begin{center}
\includegraphics[width=6.0cm,angle=270]{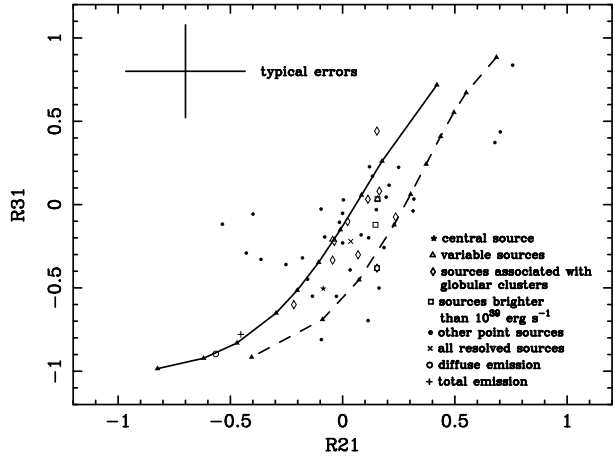}
\end{center}
\figcaption{
Hardness ratios of the X-ray point sources, defined as R21 = (M $-$ S)/(M + S)
and R31 = (H $-$ S)/(H + S), where S, M and H are the background-subtracted
counts in 0.3--1.0 keV (S), 1.0--2.0 keV (M) and 2.0--10.0 keV (H), respectively.
The solid line show the predicted hardness ratios for a power-law spectral
model with the Galactic absorption of
$2.56 \times 10^{20}$ cm$^{-2}$.
The dashed line is the same as the solid line except that the absorption is
$N_{\rm H} = 2.56 \times 10^{21}$ cm$^{-2}$.
From the top of the figure downwards, the filled triangles on the lines indicate values derived with power-law
photon indices of
0, 0.75, 1, 1.25, 1.5, 1.75, 2, 2.5, 3, and 4 respectively.
The central source, variable sources, sources associated with globular clusters,
and sources brighter than $10^{39}$ erg s$^{-1}$ are marked with a filled star,
open triangles, open diamonds and open rectangles, respectively.
The average ratios for all the resolved sources, the diffuse emission and the total
emission are marked with an X, an open circle and a cross, respectively.
\label{fig4}}
\end{figure}

\begin{figure}
\epsscale{1.0}
\begin{center}
\includegraphics[width=7.0cm,angle=270]{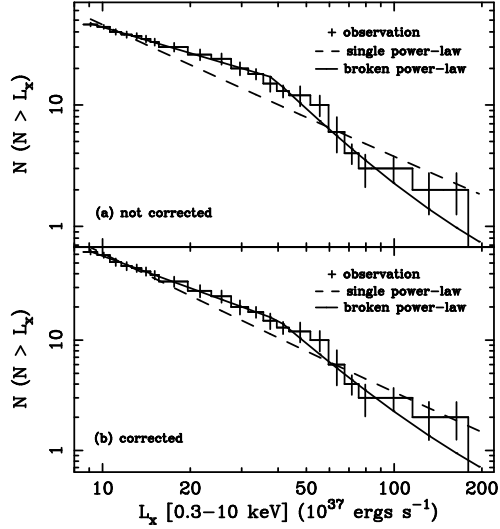}
\end{center}
\figcaption{
Cumulative X-ray luminosity functions of the off-center point sources before (a) and
after (b) the correction for sample incompleteness at the low luminosity end. The
broken power-law and power-law fits are plotted in solid and dashed lines, respectively.
\label{fig5}}
\end{figure}

\begin{figure}
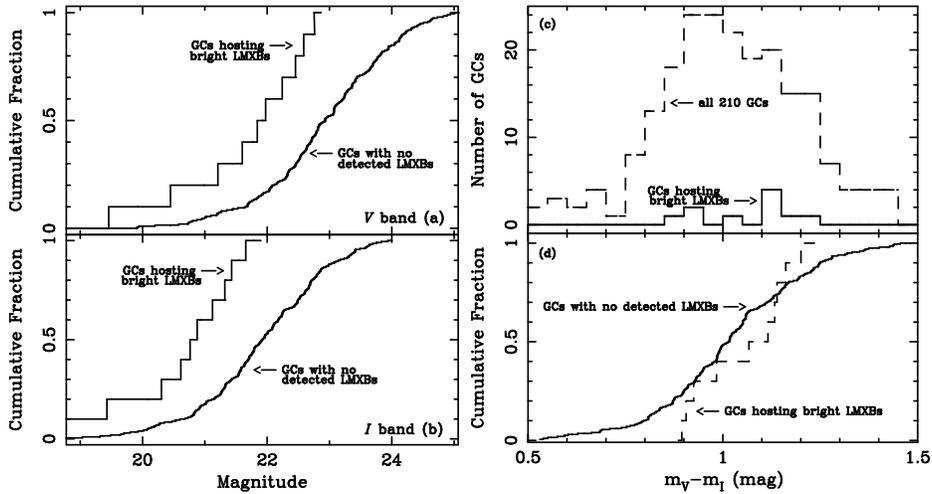

\epsscale{1.0}
\begin{center}
\includegraphics[width=6.5cm,angle=270]{f6a.ps}
\includegraphics[width=6.5cm,angle=270]{f6b.ps}
\end{center}
\figcaption{
(a)-(b): Cumulative fractional distributions of GCs that host bright LMXBs and those that do not
as a function of their $V$-band (a) and $I$-band (b) magnitudes.
(c): The optical $V-I$ color distributions of all 210 GCs (dashed line), and GCs hosting bright LMXBs (solid line). (d): Cumulative fractional distributions of GCs that host bright
LMXBs (dashed line) and those that do not (solid line) as a function of their $V-I$ color.
\label{fig6}}
\end{figure}

\begin{figure}
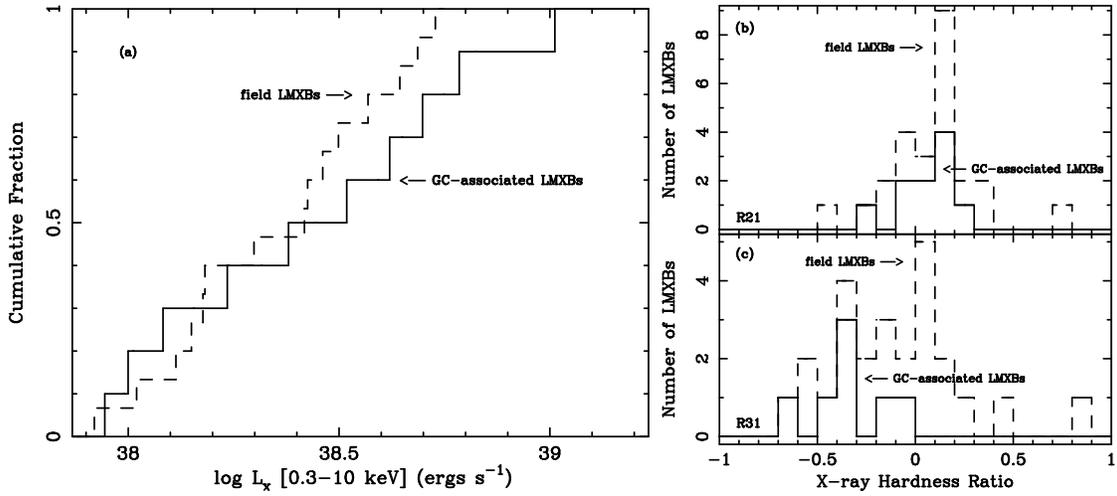

\epsscale{1.0}
\begin{center}
\includegraphics[width=6.5cm,angle=270]{f7a.ps}
\includegraphics[width=6.5cm,angle=270]{f7b.ps}
\end{center}
\figcaption{
(a): Cumulative fractional distributions of the intrinsic
X-ray luminosity of LMXBs associated with GCs (solid line), and field LMXBs (dashed line) located in the joint {\it Chandra}-{\it HST} field. (b) and (c): Histogram of the R21 (b) and R31 (c) colors for the GC-associated
(solid lines) and field (dashed lines) LMXBs, respectively.
\label{fig7}}
\end{figure}

\end{document}